\documentclass{article}
\usepackage[tbtags]{amsmath}
\usepackage{amsmath}
\usepackage[portrait, margin=1in]{geometry}
\usepackage{amssymb}
\usepackage{booktabs}
\usepackage{tikz}
\usetikzlibrary{positioning}
\usepackage{bbm}
\usepackage{authblk}
\usepackage{float}
\usepackage[toc,page]{appendix}
\usepackage{bm}
\usepackage[bottom]{footmisc}
\usepackage{tikz}
\usetikzlibrary{bayesnet}
\usetikzlibrary{arrows}
\usepackage{color}
\usepackage{graphicx}
\usepackage{caption}
\usepackage{subcaption}
\usetikzlibrary{backgrounds}
\usepackage{upgreek}
\usepackage{graphicx}
\usepackage{mathtools}
\usepackage[T1]{fontenc}
\usepackage{lipsum}   
\usepackage{setspace} 
\usepackage{algorithm}
\usepackage{algpseudocode}
\usetikzlibrary{positioning}
\usepackage[utf8]{inputenc}
\usepackage{algorithm,algcompatible,amssymb,amsmath}
\usepackage{algorithm}
\usepackage{algpseudocode}

\algnewcommand\algorithmicto{\textbf{to}}
\algnewcommand\RETURN{\State \textbf{return} }

\title{Multi-Industry Simplex 2.0 :\\ Temporally-Evolving Probabilistic Industry Classification}
\author{Maksim Papenkov$^{1,2,}$\footnote{Corresponding Author : mp3827@columbia.edu}}
\date{%
    $^1$O'Shaughnessy Asset Management\footnote{Acknowledgements: Within OSAM, I would like to specifically extend my sincere gratitude to Claire Noel, Daniel Nitiutomo, and Jai Padalkar for their invaluable contributions in bringing this project to completion.}\\%
    $^2$Columbia University, Department of Computer Science\\%
    [5ex]
    \today
}

\begin{document}

\maketitle
\vspace{-0.6cm}
\begin{abstract}  
Accurate industry classification is critical for many areas of portfolio management, yet the traditional single-industry framework of the Global Industry Classification Standard (GICS) struggles to comprehensively represent risk for highly diversified multi-sector conglomerates like Amazon. Previously, we introduced the Multi-Industry Simplex (MIS), a probabilistic extension of GICS that utilizes topic modeling, a natural language processing approach. Although our initial version, MIS-1, was able to improve upon GICS by providing multi-industry representations, it relied on an overly simple architecture that required prior knowledge about the number of industries and relied on the unrealistic assumption that industries are uncorrelated and independent over time. We improve upon this model with MIS-2, which addresses three key limitations of MIS-1 : we utilize Bayesian Non-Parametrics to automatically infer the number of industries from data, we employ Markov Updating to account for industries that change over time, and we adjust for correlated and hierarchical industries allowing for both broad and niche industries (similar to GICS). Further, we provide an out-of-sample test directly comparing MIS-2 and GICS on the basis of future correlation prediction, where we find evidence that MIS-2 provides a measurable improvement over GICS. MIS-2 provides portfolio managers with a more robust tool for industry classification, empowering them to more effectively identify and manage risk, particularly around multi-sector conglomerates in a rapidly evolving market in which new industries periodically emerge. 
\end{abstract}

\vspace{0.2cm}

\newpage

\section{Research Motivation}
\vspace{-0.2cm}

The \textbf{Multi-Industry Simplex} is a probabilistic industry classification model that we introduced in \cite{mis1_paper}, which utilizes topic modeling to represent each firm as an \textbf{industry-mixture} of the form : 
\begin{align}
    \text{firm} = \big[\text{Industry}_1 = 75\%, \text{Industry}_2 = 25\%\big]
\end{align}
Since then, we have made substantial methodological improvements that we discuss here in detail. 

\subsection{Problem}
Industry classification is critical for portfolio management, guiding stock selection and risk management. The Global Industry Classification Standard (GICS) currently dominates this space, assigning each firm to a single industry. However, this approach fails to account for the diversified nature of modern conglomerates, introducing significant \emph{misrepresentation risk}. This risk is particularly acute for market-leading multi-sector firms like Amazon, which feature prominently in many passive indices. Addressing this misrepresentation is essential for both institutional and retail investors, prompting the need for a better classification system.

\subsection{Existing Solutions (and their Limitations)}

\textbf{Global Industry Classification Standard (GICS)} has been the leading industry classification reference for over twenty years. GICS assigns each firm to exactly \emph{one} industry based on revenue and market perception factors. While this system is a reasonable low-resolution approximation of a firm, it lacks sufficient descriptive detail to appropriately represent all industry exposure risks for conglomerates. 

Apart from our natural language processing approach, several other papers have explored probabilistic industry classification methods as well \cite{blackbox1,blackbox2,blackbox3,blackbox4}, though these rely on black-box methods that introduce additional risks (see the MIS paper \cite{mis1_paper} for a detailed discussion on this). 

Finally, although the initial MIS paper (which we'll refer to as MIS-1) was able to address the single-industry constraint of GICS, it has limitations of its own that we must resolve. MIS-1 requires prior knowledge of the number of industries that exist and relies on the unrealistic assumption that industries are uncorrelated and are independent over time. We address these limitations with our improved model, MIS-2. 

\subsection{New Solution (and its Value Proposition)}

MIS-2 extends upon MIS-1 by leveraging advanced topic modelling techniques to mitigate the need for unrealistic assumptions and hyperparameter tuning. Specifically, we utilize the following : 
\vspace{-0.3cm}
\begin{itemize}
    \item \textbf{Bayesian Non-Parametrics} : a method to automatically infer the number of industries in a dataset. 
    \item \textbf{Markov Updating} : a method to dynamically update industries over time as new data arrives. 
    \item \textbf{Correlated Industries} : a method to account for similar yet distinct industries. 
    \item \textbf{Hierarchical Industries} : a method to directly model sub and super industry relationships. 
\end{itemize}
\vspace{-0.3cm}
Unlike the MIS-1 paper, which introduced the idea at a high-level, this paper is focused on the technical aspects of the model architecture. We encourage the reader to review the MIS-1 paper before reading this, as here we will assume they are already familiar with the basics of text analysis and Bayesian Learning. 

Additionally, we present an out-of-sample test comparing MIS-2 and GICS on the basis of future correlation prediction. We find evidence that MIS-2  provides a measurable improvement over GICS. We aspire to continue improving the process, and eventually provide richer backtests in future iterations.

\newpage
\section{Data Pre-Processing}
\vspace{-0.2cm}
Before delving into the model architecture, first we must prepare our data. MIS-2 utilizes text from business descriptions to identify industry membership. To mitigate the effects of noise on our signal, we construct a \textbf{keyphrase extractor} to define a subset of relevant text, which relies on the following operations : 
\vspace{-0.3cm}
\begin{enumerate}
    \item \textbf{Stemming} : reducing a word to its root (``retailer'' $\Rightarrow$ ``retail'').
    \item \textbf{Lemmatization} : replacing a word by an approximate synonym (``marketplace'' $\Rightarrow$ ``retail''). 
    \item \textbf{N-Grams} : constructing compound phrases (``e'' + ``commerce'' $\Rightarrow$ ``e-commerce'').
\end{enumerate}
\vspace{-0.3cm}
We leverage these basic tools to construct \textbf{semantic trees} that map a group of non-identical phrases to a single \textbf{semantically-unambiguous keyphrase} that summarizes the \emph{essence} of that group, while clearly corresponding to a single product or service. For example : 
\begin{figure}[h!]
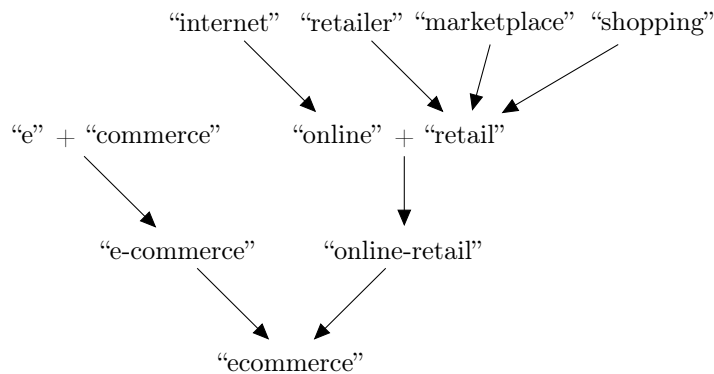

\centering
  \tikz{

  \node (ecommerce) {``ecommerce''};
  
  \node[above=of ecommerce, xshift=-1.5cm] (e-commerce) {``e-commerce''};
  \node[above=of e-commerce, xshift=-1.5cm] (plus_left) {+};
  \node[left=of plus_left, xshift=+1.1cm, yshift=+0.05cm] (e) {``e''};
  \node[right=of plus_left, xshift=-1.1cm, yshift=+0.05cm] (commerce) {``commerce''};
  
  \node[above=of ecommerce, xshift=+1.5cm] (online_retail) {``online-retail''};
  \node[above=of online_retail] (plus_right) {+};
  \node[left=of plus_right, xshift=+1.1cm, yshift=+0.05cm] (online) {``online''};
  \node[right=of plus_right, xshift=-1.1cm, yshift=+0.05cm] (retail) {``retail''};
  \node[above=of online, xshift=-1.5cm] (internet) {``internet''};
  \node[above=of retail, xshift=-1.5cm] (retailer) {``retailer''};
  \node[above=of retail, xshift=0.4cm, yshift=-0.05cm] (marketplace) {``marketplace''};
  \node[above=of retail, xshift=+2.5cm, yshift=-0.05cm] (shopping) {``shopping''};

  \draw[->, shorten >=2pt] (e-commerce)  -- (ecommerce);
  \draw[->, shorten >=2pt] (plus_left)  -- (e-commerce);
  \draw[->, shorten >=2pt] (online_retail)   -- (ecommerce);
  \draw[->, shorten >=2pt] (plus_right) -- (online_retail);
  \draw[->, shorten >=2pt] (internet) -- (online);
  \draw[->, shorten >=2pt] (retailer) -- (retail);
  \draw[->, shorten >=2pt] (marketplace) -- (retail);
  \draw[->, shorten >=2pt] (shopping) -- (retail);

 }
 \caption{Partial Semantic-Tree for Keyphrase = ``ecommerce''}
\end{figure}\label{semantic_tree}
\vspace{-0.4cm}

As semantics is inherently subjective, the construction of such trees requires careful discretion and domain expertise. For our particular implementation of MIS-2, we construct over 300 semantic trees that map over 9000 n-grams to semantically-unambiguous keyphrases. As an illustrative example, consider Pitchbook's 2023 business description for Amazon along with industry-related phrases concatenated to the end of the document. Here we show our keyphrase extractor in action : 
\begin{figure}[h!]
    \centering
    \includegraphics[width=16.5cm]{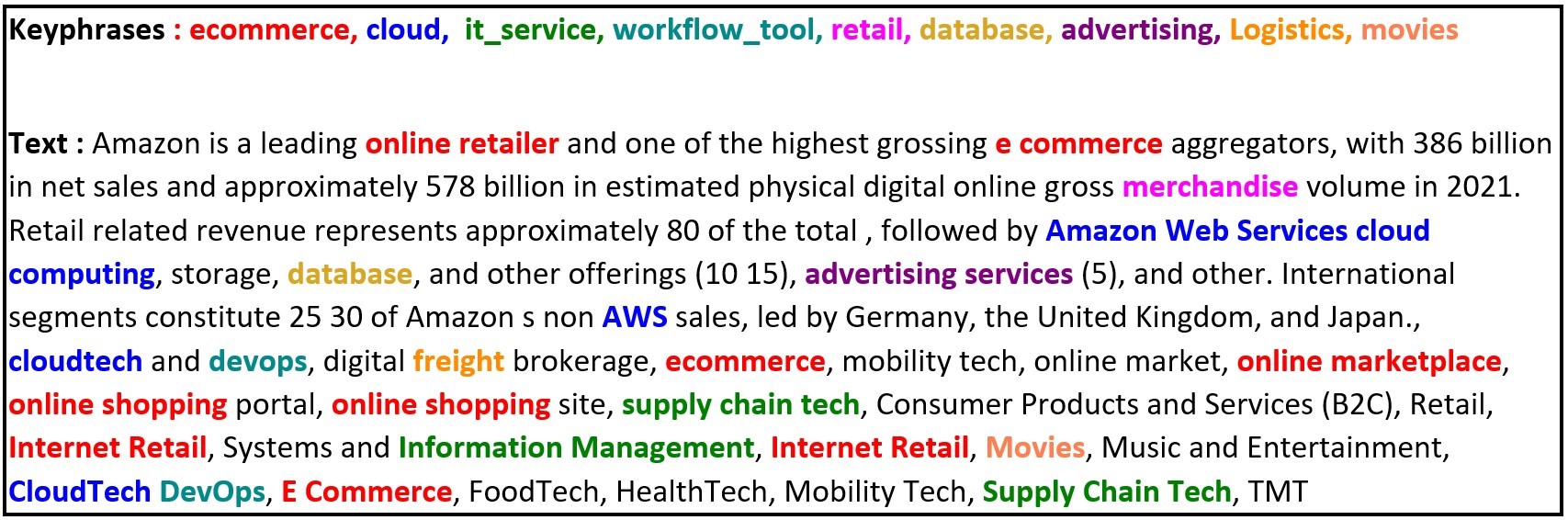}
    \caption{MIS-2 Keyphrase Extraction for Amazon}
\end{figure}

\vspace{-0.5cm}
We replace highlighted n-grams with their associated semantically-unambiguous keyphrases (each with its own color) and discard all remaining text. Our final data object is thus a \textbf{bag-of-words} of the form : 
\begin{align}
    \text{Amazon} = \{\text{ecommerce, ecommerce, retail, cloud, cloud, ...}\}
\end{align} 
This is the input format necessary to fit a topic model. 

\newpage
\section{Topic Modeling for Probabilistic Industry Classification}

A \textbf{topic model} identifies clusters of frequently co-occurring words within a corpus. For sufficiently clean data, these clusters are often human-interpretable and share a common topic (hence the name). Since we pre-process our text such that our vocabulary only includes semantically-unambiguous phrases relating to products and services, we define each word-cluster as an \textbf{MIS-industry} (e.g. a cluster including ``computer-vision'', ``machine-learning'', and ``natural-language-processing'' implies \emph{artificial intelligence}). 

Once a topic model is trained, we can represent each input document as a probability distribution over topics, which in this case each correspond to an industry. Thus, we can represent each business description as an \emph{industry-mixture}, allowing us to efficiently and systematically identify industry relevance for firms. 

For example, consider the text from Figure 2, which would be processed as : 
\begin{figure}[h!]
\centering
\begin{tikzpicture}
 
    \node[rectangle,draw, align=left] (business_description) at (0,0) {``Amazon is a leading\\ online retailer and one \\of the highest grossing\\e commerce ... ''};
    
    \node[rectangle,draw, right=of business_description, align=left] (freq_words) {``ecommerce'' x 9\\``cloud'' x 5\\ ``movie'' x 1\\
    ...};

    \node[rectangle,draw, right=of freq_words, align=left] (raw_topics) {\emph{ecommerce} ($40\%$)\\\emph{cloud} ($30\%$)\\ \emph{movies} ($20\%$)\\
    ...};

        \node[above=of business_description, yshift=-0.9cm] {Raw Text};
        \node[above=of freq_words, yshift=-0.9cm] {Bag-of-Words};
        \node[above=of raw_topics, yshift=-0.9cm] {Industry-Mixture};

    \draw[->, shorten >=3pt] (business_description) -- (freq_words);
    \draw[->, shorten >=3pt] (freq_words) -- (raw_topics);

\end{tikzpicture}
 \caption{MIS-Industry Probabilities for Amazon}
\end{figure}
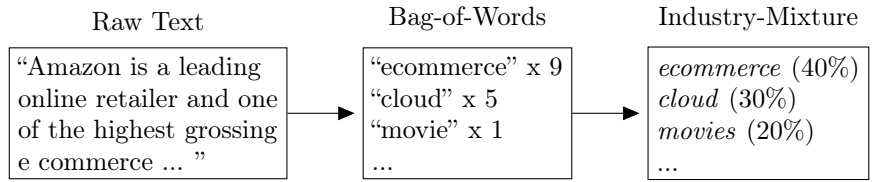

\vspace{-0.3cm}
The only knowledge that the topic model has about a firm is contained within the input text, so careful data sourcing and pre-processing is imperative to guarantee a reliable output. We'll now attempt to explain and demystify this process by making it mathematically precise. 
\vspace{-0.3cm}

\subsection{Components of a Topic Model}

First, we define our data as : 
\vspace{-0.3cm}
\begin{itemize}
    \item $\mathbf X = \{\mathbf x_1, \dots, \mathbf x_M\}$ : a \textbf{corpus} containing $M$-many business descriptions. 
    \item $\mathbf x_m$ : the $m$-th \textbf{business description}, represented as a \emph{bag-of-words} containing $N_m$-many keyphrases. 
    \item $\mathbf x_{m,n}$ :  the $n$-th \textbf{keyphrase} of the $m$-th business description.  
\end{itemize}
\vspace{-0.3cm}
Then, we fit a topic model with $K$-many topics, such that each is a distribution over a vocabulary of $V$-many distinct keyphrases. For the remainder of this paper, we'll refer to these topics as \emph{MIS-industries} within the context of our application. In practice, we ignore probabilities below some threshold, such that each MIS-industry corresponds to only a subset of the keyphrases. The notation $\triangle^V$ represents a \textbf{simplex}, which is a discrete probability distribution over $V$-many categories. 
\vspace{-0.3cm}
\begin{itemize}
    \item $\boldsymbol\phi_k \in \triangle^V$ : the $k$-th \textbf{MIS-Industry}, which is a distribution over $V$-many keyphrases.
\end{itemize}
\vspace{-0.3cm}
Finally, for each input firm with business description $\mathbf x_m$ we estimate an \emph{industry-mixture} over our $K$-many MIS-industries. For these we also ignore small probabilities, such that each firm is only associated with a small subset of the most probable industries. 
\vspace{-0.3cm}
\begin{itemize}
    \item $\boldsymbol\theta_m \in \triangle^K$ : the \textbf{industry-mixture} corresponding to to the $m$-th firm's business description. 
\end{itemize}
\vspace{-0.3cm}
One additional technical detail that is relevant for parameter inference is that each individual keyphrase in each individual bag-of-words is assigned to a single MIS-industry via an industry-index. This latent variable is essential for inferring $\boldsymbol\phi_k$ and $\boldsymbol\theta_m$, as we'll see in a later section.
\vspace{-0.3cm}
\begin{itemize}
    \item $\mathbf z_{m,n}$ : the \textbf{industry-index} for keyphrase $\mathbf x_{m,n}$ (such that $\mathbf z_{m,n} \in \{1, 2, \dots, K\})$. 
\end{itemize}
\vspace{-0.3cm}
We will consider multiple topic model architectures; the natural starting point is \emph{Latent Dirichlet Allocation}. 

\newpage
\subsection{MIS-1 Architecture (Latent Dirichlet Allocation)}

\textbf{Latent Dirichlet Allocation (LDA)} \cite{blei_LDA} is the fundamental starting point for probabilistic topic modeling, and is the primary architecture that we used for MIS-1 \cite{mis1_paper}. Note - for the remainder of this paper, we assume the reader is familiar with Bayesian learning and graphical modeling. 

In addition to the components from the previous section, LDA has two hyperparameters on its prior :  
\vspace{-0.3cm}
\begin{itemize}
    \item $\boldsymbol 
    \alpha$ : a control on $\boldsymbol\phi_k$ that influences the number of keyphrases associated with each MIS-industry.
    \item $\boldsymbol\beta$ : a control on $\boldsymbol\theta_m$ that influences the number of MIS-industries associated with each industry-mixture.
\end{itemize}
\vspace{-0.3cm}
We relate all of the parameters together with the following generative process : 
\begin{align}
    \mathbf x_{m, n} &\sim \text{Categorical}_V\big(\mathbf x_{m,n} \mid \boldsymbol\phi_{z_{m,m}}\big)
    \\
    \mathbf z_{m,n} &\sim \text{Categorical}_K\big(\mathbf z_{m,n} \mid \boldsymbol\theta_m\big)
    \\
    \boldsymbol\theta_m &\sim \mathbb P\big(\text{MIS-Industry} \mid \text{Firm}_m\big) = \text{Dirichlet}_K\big(\boldsymbol\theta_m\mid \boldsymbol\beta\big)
    \\
    \boldsymbol\phi_{k} &\sim \mathbb P\big(\text{Keyphrase}\mid \text{MIS-Industry}_k\big)=\text{Dirichlet}_V\big(\boldsymbol\phi_{k}\mid \boldsymbol\alpha\big)
\end{align}
This generative process is summarized in the following graphical model : 
\begin{figure}[h!]
\centering
  \begin{tikzpicture}
 \node[obs                                  ] (x)     {$\mathbf x_{m,n}$};
 \node[latent, right=of x    , xshift=+0.3cm] (z)     {$\mathbf z_{m,n}$}; 
 \node[latent, left=of x     , xshift=-0.3cm] (phi)   {$\boldsymbol\phi_k$};
 \node[const ,left=of phi   , xshift=-0.1cm] (alpha) {$\boldsymbol\alpha$};
 \node[latent, right=of z    , xshift=+0.3cm] (theta) {$\boldsymbol\theta_m$};
 \node[const, right=of theta, xshift=+0.1cm] (beta)  {$\boldsymbol\beta$};

 \node[above=of phi  , align=center] (phi_label)   {Keyphrase-Mixture\\for MIS-Industry-k};
 \node[above=of theta, align=center] (theta_label) {Industry-Mixture\\for Firm-m};
 \node[below=of x    , align=center] (x_label)     {Keyphrase-n\\ for Firm-m};
 \node[below=of z    , align=center] (z_label)     {Index-n\\ for Firm-m};
 \node[left=of alpha , align=center, xshift=0.9cm] (alpha_label) {Keyphrase-Mixture\\Hyperparameter};
 \node[right=of beta , align=center, xshift=-0.9cm] (beta_label) {Industry-Mixture\\Hyperparameter};
 
 \plate [inner sep=.4cm,yshift=.2cm] {plate1} {(x)(z)} {$N_m$}; 
 \plate [inner sep=.6cm,yshift=.2cm] {plate2} {(x)(z)(theta)} {$M$};
 \plate [inner sep=.3cm,yshift=.2cm] {plate3} {(phi)} {$K$};

 \draw[->, shorten >=3pt] (z) -- (x);
 \draw[->, shorten >=3pt] (phi) -- (x);
 \draw[->, shorten >=1pt] (alpha) -- (phi);
 \draw[->, shorten >=3pt] (theta) -- (z);
 \draw[->, shorten >=3pt] (beta) -- (theta);

 \end{tikzpicture}
 \caption{Latent Dirichlet Allocation}
\end{figure}
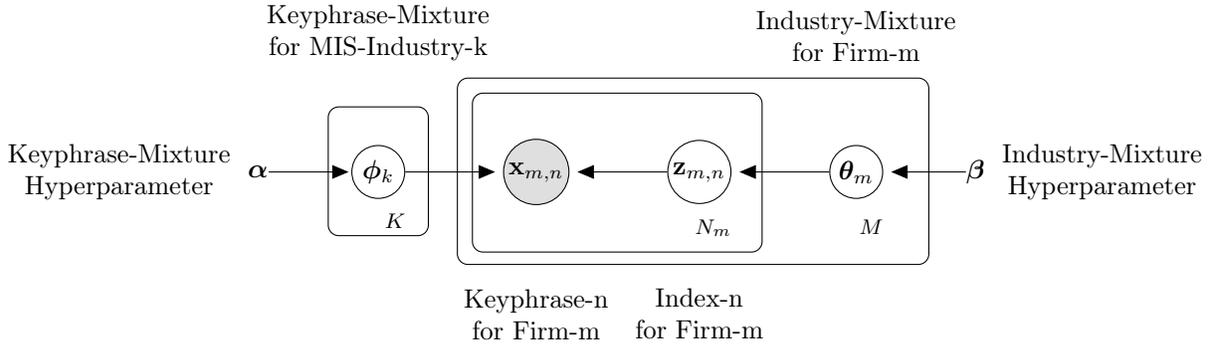

\vspace{-0.3cm}
Posterior inference can be performed via Gibbs Sampling \cite{newman2009} or Variational Inference \cite{blei_LDA} (for details on the LDA Gibbs Sampler, see the MIS-1 paper). Without getting into the numerical details of inference, we can obtain useful intuitions about the posterior by looking closely at the joint log-likelihood of LDA : 
\begin{equation}
    \log \mathbb P\big(\boldsymbol\theta_{1:M}, \boldsymbol\phi_{1:K} \;; \mathbf x_{1:M}\big) = \sum_{k=1}^K \log\mathbb P\big(\boldsymbol\phi_k\big) +
    \sum_{m=1}^M\bigg[\log \mathbb P\big(\boldsymbol\theta_m\big) + \sum_{i=1}^{N_m}
    \big(\log\boldsymbol\theta_{m, z_{m,i}} + \log \boldsymbol\phi_{z_{m,i},w_{m,i}}\big)
    \bigg]
\end{equation}
Notice that there are two mechanisms that maximize this function : 
\vspace{-0.3cm}
\begin{itemize}
    \item For the $m$-th firm, select the MIS-industries that maximize probability $\mathbb P(\boldsymbol\theta_m\big)$. 
    \item For the $k$-th MIS-industry, select keyphrases that maximize probability $\mathbb P\big(\boldsymbol\phi_k\big)$.
\end{itemize}
\vspace{-0.3cm}
Further, since each $\boldsymbol\theta_m$ and $\boldsymbol\phi_k$ is a simplex variable that necessarily sums to one, this objective encourages the emergence of highly concentrated distributions, such that each firm is associated with few MIS-industries, and each MIS-industry is associated with few keyphrases. Finally, notice that these are competing goals - the fewer MIS-industries per firm, the more pressure exists on those MIS-industries to cover all of the keyphrases in the firm's business description. Thus, our posterior is an optimally compact representation of each firm such that its associated MIS-industries are only those most strongly justified by the data. 

LDA requires us to know $K$ ahead of time, which is difficult to tune in practice, and assumes that industries are independent both cross-sectionally and over time. We'll now turn to more advanced topic model architectures to address these limitations. 

\newpage
\subsection{Improvement 1 : Automatically Inferring the Number of Industries}
\emph{LDA Limitation} : For LDA it is unclear how to reasonably select $K$, which represents the total number of MIS-Industries, though we can automate this selection with \emph{Bayesian Non-Parametrics}. 

\textbf{Bayesian Non-Parametrics} \cite{bayesian_non_parametrics} is a modeling framework that allows us to ``discover'' an optimal $K$ during inference, such that there are precisely as many dimensions as can be supported by the dataset. Though the underlying theory of non-parametrics can be esoteric, we provide three simple perspectives :
\vspace{-0.3cm}
\begin{itemize}
    \item $K=\infty$ : in an \emph{abstract mathematical} setting, we posit that there exist infinitely-many \emph{potential} industries, but for a finite dataset we only observe a \emph{finite subset} of them. This perspective provides a basis for a robust theory that is beyond the scope of this paper. 
    \item $K=\;?$ : in a \emph{computational} setting, we treat the number industries as a latent random variable that can be \emph{inferred} from a data sample. This perspective enables us to leverage statistical methods in order to design algorithms that can run in finite-time to estimate $K$ from a dataset.
    \item $K_t \sim \{K\}_{1,2,\dots,T}$ : in a \emph{phenomenological} setting, we treat the number of industries at time-t as a stochastic process that is non-constant over time. This perspective allows us to evolve the number of industries within a changing market, where new industries emerge and obsolete industries disappear. 
\end{itemize}
\vspace{-0.3cm}
Without being too mathematically formal, we loosely define a \emph{Dirichlet Process} as a discrete distribution over infinitely-many categories. We can then extend LDA to a non-parametric setting as a \textbf{Hierarchical Dirichlet Process} \cite{HDP}, which is defined by the following generative model as $K\rightarrow\infty$  : 
\begin{align}
    \mathbf x_{m, n} &\sim \text{Categorical}_V\big(\mathbf x_{m,n} \mid \boldsymbol\phi_{z_{m,m}}\big)
    \\
    \mathbf z_{m,n} &\sim \text{Categorical}_K\big(\mathbf z_{m,n} \mid \boldsymbol\theta_m\big)
    \\
    \boldsymbol\theta_m &\sim \mathbb P\big(\text{MIS-Industry} \mid \text{Firm}_m\big) = \text{Dirichlet}_K\big(\boldsymbol\theta_m\mid \boldsymbol\beta\eta\big)
    \\
    \boldsymbol\phi_{k} &\sim \mathbb P\big(\text{Keyphrase}\mid \text{MIS-Industry}_k\big)=\text{Dirichlet}_V\big(\boldsymbol\phi_{k}\mid \boldsymbol\alpha\big)
    \\
    \boldsymbol\eta &\sim \text{Dirichlet}_K\big(\boldsymbol\eta\mid \gamma\big)
\end{align}
This is very similar to LDA, though with an additional variable $\boldsymbol\eta$ that is associated with a global set for topics, which is controlled by one extra hyperparameter : 
\vspace{-0.3cm}
\begin{itemize}
    \item $\gamma$ : a control on the final estimated $K$ (a higher $\gamma$ leads to a higher final $K$)
\end{itemize}
\vspace{-0.3cm}
At a high-level, $\boldsymbol\theta_m$ represents the distribution of MIS-industries for a single firm, while $\boldsymbol\eta$ represents the distributions of MIS-industries over an entire universe. Intuitively, broad industries like banking include many firms, while niche industries like lumber milling only include a few. This process is summarized in the following graphical model, note the similarities to LDA. 
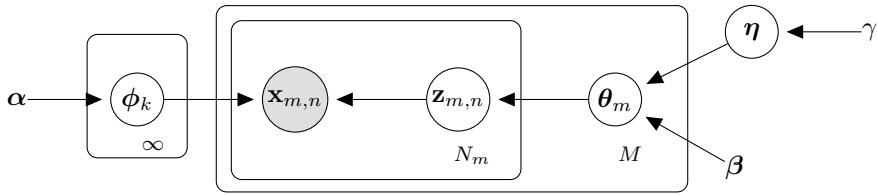
\begin{figure}[h!]
\centering
  \begin{tikzpicture}
 \node[obs                                  ] (x)     {$\mathbf x_{m,n}$};
 \node[latent, right=of x    , xshift=+0.3cm] (z)     {$\mathbf z_{m,n}$}; 
 \node[latent, left=of x     , xshift=-0.3cm] (phi)   {$\boldsymbol\phi_k$};
 \node[const, left=of phi    , xshift=-0.1cm] (alpha) {$\boldsymbol\alpha$};
 \node[latent, right=of z    , xshift=+0.3cm] (theta) {$\boldsymbol\theta_m$};
 \node[latent, right=of theta, xshift=+0.1cm, yshift=+0.9cm] (eta)  {$\boldsymbol\eta$};
 \node[const, right=of theta, xshift=+0.1cm, yshift=-0.9cm] (beta)  {$\boldsymbol\beta$};
 \node[const, right=of eta  , xshift=+0.1cm] (gamma)  {$\gamma$};
 
 \plate [inner sep=.4cm,yshift=.2cm] {plate1} {(x)(z)} {$N_m$}; 
 \plate [inner sep=.6cm,yshift=.2cm] {plate2} {(x)(z)(theta)} {$M$};
 \plate [inner sep=.3cm,yshift=.2cm] {plate3} {(phi)} {$\infty$};

 \draw[->, shorten >=3pt] (z) -- (x);
 \draw[->, shorten >=3pt] (phi) -- (x);
 \draw[->, shorten >=3pt] (alpha) -- (phi);
 \draw[->, shorten >=3pt] (theta) -- (z);
 \draw[->, shorten >=3pt] (eta) -- (theta);
 \draw[->, shorten >=3pt] (gamma) -- (eta);
 \draw[->, shorten >=3pt] (beta) -- (theta);

 \end{tikzpicture}
 \caption{Hierarchical Dirichlet Process (also called an ``Infinite Topic Model'')}
\end{figure}
\vspace{-0.3cm}

While there exist many technical subtleties to consider when implementing an HDP in practice, a discussion of the rich literature on this topic is beyond the scope of this paper (though we encourage the reader to review \cite{hdp_inference, hdp_gibbs} for more detail). For our purpose however, we simply note that an HDP is an extension of LDA that estimates $K$ during the parameter inference process. Though $K$ is infinite \emph{in theory}, for a trained topic model it's simply a finite integer.

\newpage
\subsection{Improvement 2 : Modeling Industry Correlations and Hierarchies}

\emph{LDA Limitation} : Another critical limitation of LDA is that by representing each industry-mixture as a Dirichlet variable we embed the implicit assumption that all industry categories are \emph{independent}. This is unrealistic, as many pairs of industries are correlated - such as \emph{artificial-intelligence} which often co-occurs with \emph{robotics}, and \emph{oil-drilling} which often co-occurs with \emph{petrochemicals}.

A compelling alternative is the \textbf{Correlated Topic Model (CTM)} \cite{blei_CTM}, which in most regards is identical to LDA apart from one key modification - it replaces the single-variable Dirichlet distribution on $\boldsymbol\theta_m$ with the two-variable \textbf{Logistic Normal} distribution. The Logistic Normal distribution involves sampling a Multivariate Gaussian variable and transforming it into a simplex via the following approach : 
\begin{align}\label{logistic_normal}
    \boldsymbol\theta \sim \text{LogisticNormal}\big(\boldsymbol\mu, \boldsymbol\Sigma\big) \iff 
    \boldsymbol\eta \sim \text {Normal}\big(\boldsymbol\mu, \boldsymbol\Sigma\big) \text{ } ;\text { } 
    \boldsymbol\theta = \frac{\exp \boldsymbol\eta_i}{\sum \boldsymbol\exp \eta_i} 
\end{align}
Thus, the generative process for CTM is : 
\begin{align}
    \mathbf x_{m, n} &\sim \text{Categorical}_V\big(\mathbf x_{m,n} \mid \boldsymbol\phi_{z_{m,m}}\big)
    \\
    \mathbf z_{m,n} &\sim \text{Categorical}_K\big(\mathbf z_{m,n} \mid \boldsymbol\theta_m\big)
    \\
    \boldsymbol\theta_m &\sim \mathbb P\big(\text{MIS-Industry} \mid \text{Firm}_m\big) = \text{LogisticNormal}_K\big(\boldsymbol\theta_m\mid \boldsymbol\mu, \boldsymbol\Sigma\big)
    \\
    \boldsymbol\phi_{k} &\sim \mathbb P\big(\text{Keyphrase}\mid \text{MIS-Industry}_k\big)=\text{Dirichlet}_V\big(\boldsymbol\phi_{k}\mid \boldsymbol\alpha\big)
\end{align}
In plain English, the CTM's $\boldsymbol\Sigma$ parameter represents the correlations between pairs of MIS-industries, which allows us to directly model \emph{similar, yet distinct} MIS-industries within a single model. Such a framework is much more representative of the real world, as the orthogonality of the LDA Dirichlet can lead to cases of \emph{greedy inference} in which a firm only ends up associated with one of its two industries rather than both in the posterior. We summarize the CTM in the following graphical model : 
\begin{figure}[h!]
\centering
  \begin{tikzpicture}
 \node[obs                                  ] (x)     {$\mathbf x_{m,n}$};
 \node[latent, right=of x    , xshift=+0.3cm] (z)     {$\mathbf z_{m,n}$}; 
 \node[latent, left=of x     , xshift=-0.3cm] (phi)   {$\boldsymbol\phi_k$};
 \node[const, left=of phi   , xshift=-0.1cm] (alpha) {$\boldsymbol\alpha$};
 \node[latent, right=of z    , xshift=+0.3cm] (theta) {$\boldsymbol\theta_m$};
 \node[const, right=of theta, xshift=+0.1cm, yshift=+0.8cm] (mu)  {$\boldsymbol\mu$};
 \node[const, right=of theta, xshift=+0.1cm, yshift=-0.8cm] (sigma)  {$\boldsymbol\Sigma$};
 
 \plate [inner sep=.4cm,yshift=.2cm] {plate1} {(x)(z)} {$N_m$}; 
 \plate [inner sep=.6cm,yshift=.2cm] {plate2} {(x)(z)(theta)} {$M$};
 \plate [inner sep=.3cm,yshift=.2cm] {plate3} {(phi)} {$K$};

 \draw[->, shorten >=3pt] (z) -- (x);
 \draw[->, shorten >=3pt] (phi) -- (x);
 \draw[->, shorten >=1pt] (alpha) -- (phi);
 \draw[->, shorten >=3pt] (theta) -- (z);
 \draw[->, shorten >=3pt] (mu) -- (theta);
 \draw[->, shorten >=3pt] (sigma) -- (theta);

 \end{tikzpicture}
 \caption{Correlated Topic Model}
\end{figure}
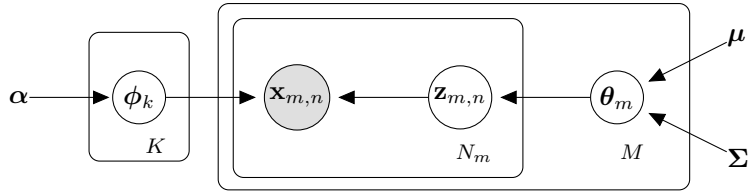

\vspace{-0.3cm}
Despite the intuitive appeal of the CTM, we however ran into numerical issues during parameter inference that prevented us from using this particular architecture in our final version of MIS-2. For our implementation of MIS-2, we chose to only use Gibbs Sampling rather than Variational Inference due to accuracy compromises associated with the mean-field assumption, so we only tested a Gibbs Sampler for CTM. The primary numerical appeal of LDA is that the Dirichlet and Categorical distributions are conjugate, which makes each individual step of Gibbs Sampling or Variational Inference more efficient. By substituting the Logistic Normal into the model, we lose conjugacy and this leads to significantly slower inference (see \cite{ctm_sampler}) This method was very slow and unstable and we thus elected to pursue another route.

We also similarly considered \textbf{Hierarchical LDA (HLDA)} \cite{blei_HTM}, which produces a topic tree, with sub and super-topics, though we also ran into numerical instability that prevented us from utilizing this in practice. 

Although we do not use the CTM or HLDA architectures directly, these two brilliant papers have had a profound influence on MIS-2. In reality, some pairs of industries are correlated, and others are subsets of larger categories - and thus we \emph{must} account for this at some point in our process. Rather than address this in the model architecture however, we instead resolve this in post-processing (Section 4).

\newpage
\subsection{Improvement 3 : Evolving Industries over Time}

\emph{LDA Limitation} : Although industries change over time, LDA assumes independence over different annual datasets. To construct an evolving model, we leverage the spirit of the Dynamic Topic Model. 

The \textbf{Dynamic Topic Model} \cite{blei_DTM} is a time-series extension of LDA in which we add a stochastic process over our parameters to allow the model to temporally evolve. In the original paper, this evolution is represented with a Kalman Filter \cite{kalman_filter}, which is a Markov Process (in which time $t$ only depends on time $t-1$) with Gaussian noise for some fixed variance $\sigma^2\in\mathbb R_+$ : 
\begin{align}
    \boldsymbol\alpha_t \sim \text{Normal}\big(\boldsymbol\alpha_t \mid \boldsymbol\alpha_{t-1}, \sigma^2 \mathbf I\big)
\end{align}
To ensure that our new variable is also a well-defined simplex, we must apply the same normalizing transform as in Equation \ref{logistic_normal}. The full generative process is summarized as : 
\begin{align}
    \mathbf x_{m, n} &\sim \text{Categorical}_V\big(\mathbf x_{m,n} \mid \boldsymbol\phi_{z_{m,m}}\big)
    \\
    \mathbf z_{m,n} &\sim \text{Categorical}_K\big(\mathbf z_{m,n} \mid \boldsymbol\theta_m\big)
    \\
    \boldsymbol\theta_m &\sim \mathbb P\big(\text{MIS-Industry} \mid \text{Firm}_m\big) = \text{Dirichlet}_K\big(\boldsymbol\theta_m\mid \boldsymbol\beta_t\big)
    \\
    \boldsymbol\phi_{k} &\sim \mathbb P\big(\text{Keyphrase}\mid \text{MIS-Industry}_k\big)=\text{Dirichlet}_V\big(\boldsymbol\phi_{k}\mid \boldsymbol\alpha_t\big)
    \\
    \boldsymbol\alpha_t &\sim \text{Normal}\big(\boldsymbol\alpha_t \mid \boldsymbol\alpha_{t-1}, \sigma^2_\alpha \mathbf I\big)
    \\
    \boldsymbol\beta_t &\sim \text{Normal}\big(\boldsymbol\beta_t \mid \boldsymbol\beta_{t-1}, \sigma^2_\beta \mathbf I\big)
\end{align}
The temporal evolution of $\boldsymbol\alpha$ and $\boldsymbol\beta$ is visually represented in the following graphical model (suppressing the $t$ subscripts within the plates for readability): 
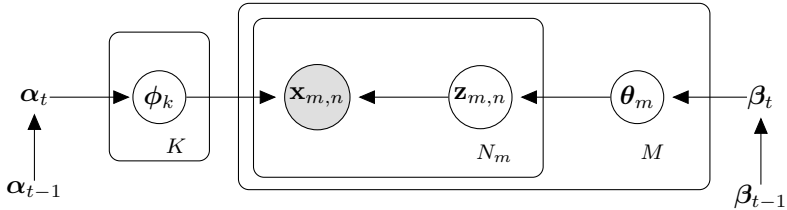
\begin{figure}[h!]
\centering
  \begin{tikzpicture}
 \node[obs                                  ] (x)     {$\mathbf x_{m,n}$};
 \node[latent, right=of x    , xshift=+0.3cm] (z)     {$\mathbf z_{m,n}$}; 
 \node[latent, left=of x     , xshift=-0.3cm] (phi)   {$\boldsymbol\phi_k$};
 \node[const ,left=of phi   , xshift=-0.1cm] (alpha) {$\boldsymbol\alpha_{t}$};
  \node[const ,below=of alpha   , xshift=-0.0cm] (alpha2) {$\boldsymbol\alpha_{t-1}$};
 \node[latent, right=of z    , xshift=+0.3cm] (theta) {$\boldsymbol\theta_m$};
 \node[const, right=of theta, xshift=+0.1cm] (beta)  {$\boldsymbol\beta_{t}$};
  \node[const, below=of beta, xshift=+0.0cm] (beta2)  {$\boldsymbol\beta_{t-1}$};
 
 \plate [inner sep=.4cm,yshift=.2cm] {plate1} {(x)(z)} {$N_m$}; 
 \plate [inner sep=.6cm,yshift=.2cm] {plate2} {(x)(z)(theta)} {$M$};
 \plate [inner sep=.3cm,yshift=.2cm] {plate3} {(phi)} {$K$};

 \draw[->, shorten >=3pt] (z) -- (x);
 \draw[->, shorten >=3pt] (phi) -- (x);
 \draw[->, shorten >=1pt] (alpha) -- (phi);
 \draw[->, shorten >=3pt] (theta) -- (z);
 \draw[->, shorten >=3pt] (beta) -- (theta);
 \draw[->, shorten >=3pt] (beta2) -- (beta);
 \draw[->, shorten >=3pt] (alpha2) -- (alpha);

 \end{tikzpicture}
 \caption{Dynamic Topic Model}
\end{figure}

\vspace{-0.3cm}
While this architecture effectively captures the temporal changes that we care about, it has one practical limitation that's a particular issue for backtesting - it models all time-slices simultaneously, implicitly introducing \emph{look-ahead bias}. This is an issue in which the parameters at time $t_1$ are optimized given information for some $t_2$ in the future such that $t_1 < t_2$. We thus require a simple modification of this approach. 

Rather than fit a single DTM to all data, we instead fit a sequence of LDA models that utilize a similar Markov parameter process, such that we strictly avoid looking into the future during inference. For this we require more detailed notation : at time $t$ let $\boldsymbol\alpha_{t,k}^0$ denote the prior parameter, and let $\boldsymbol\alpha_{t,k}^*$ denote the posterior parameter. We then modify our LDA generative process prior at time $t$ to be a function of the posterior of LDA model trained at time $t-1$ : 
\begin{align}
    \boldsymbol\phi_{t,k} \sim \text{Dirichlet}_V\big(\boldsymbol\phi_{t,k} \mid \boldsymbol\alpha_{t-1,k}^*\big)
\end{align}
In practice, assuming we fit annual MIS models, we simply use last-year's posterior as this-year's prior. This creates a path-dependent process in which we simply propagate information forward and allow our MIS-industries to evolve year-over-year. To avoid overfitting to the past, we can also apply a simple transform like $f(x) = x^{0.5}$ to $\boldsymbol\alpha^*_{t-1,k}$ to make it closer to a uniform distribution, which dilutes the prior and allows the new data to have a stronger influence on the posterior. 

We are now equipped to apply the lessons we have learned from these various topic model architectures to create a single ensemble model that addresses everything we care about. 

\newpage
\subsection{MIS-2 Architecture (Ensembled Topic Model)}

For MIS-2, we create an ensemble architecture that incorporates elements of each of the topic models that we have discussed. At a high-level, we perform the following process :
\vspace{-0.3cm}
\begin{enumerate}
    \item For $\text{year}=1$, we independently fit $S$-many HDP models to ``discover'' MIS-industries. These models are fit with uninformative priors, identical hyperparameters, but different random seeds. We then ensemble together all posteriors by retaining MIS-industries that appear in sufficiently many ensemble members, which mitigates the risk of spurious identification (inspired by \cite{ensemble_LDA}). 
    \item For $\text{year}=1$, we fit an LDA model using the Empirical Bayes prior that we constructed from our HDP ensemble, where $K$ equals the number of all non-trivial MIS-industries discovered in the first step. 
    \item For $\text{year}=t$, we fit an LDA model using a strong prior which is based on the posterior of the previous year's model, allowing us to temporally evolve our existing MIS-industries. 
    \item For $\text{year}=T$ (the current year), we then adjust the estimated industry-mixtures from the final LDA model to account for correlated and hierarchical industry relationships to obtain our final MIS-industry relevances for each firm (more on this in the next section). 
\end{enumerate}
\vspace{-0.3cm}
This process is summarized in the following flow chart : 
\begin{figure}[h!]
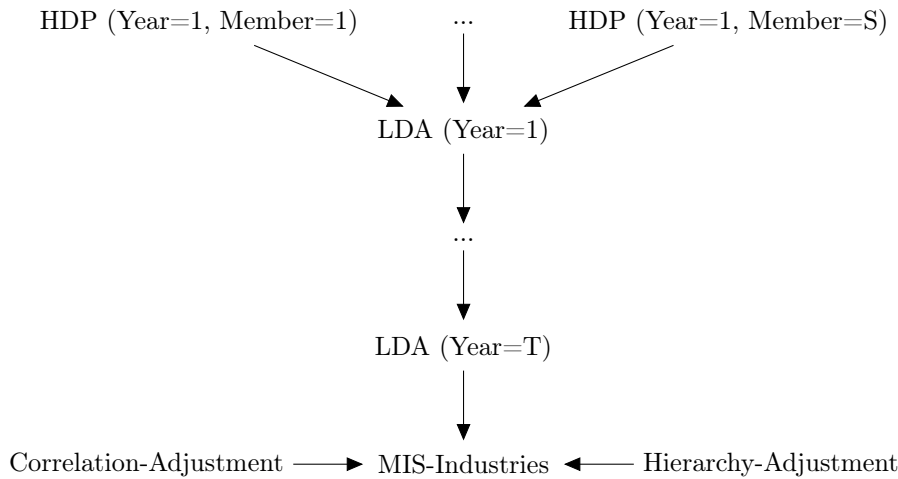

    \centering
    \tikz{
        \node(HDP2) {...};
        \node[left=of HDP2] (HDP1) {HDP (Year=1, Member=1)};
        \node[right=of HDP2] (HDP3) {HDP (Year=1, Member=S)};
        \node[below=of HDP2](LDA1) {LDA (Year=1)};
        \node[below=of LDA1] (LDAmid) {...};
        \node[below=of LDAmid] (LDAT) {LDA (Year=T)};
        \node[below=of LDAT] (mis) {MIS-Industries};
        \node[left=of mis] (corr) {Correlation-Adjustment};
        \node[right=of mis] (hierarchy) {Hierarchy-Adjustment};

        \draw[->, shorten >=2pt] (HDP1)  -- (LDA1);
        \draw[->, shorten >=2pt] (HDP2)  -- (LDA1);
        \draw[->, shorten >=2pt] (HDP3)  -- (LDA1);
        \draw[->, shorten >=2pt] (LDA1)  -- (LDAmid);
        \draw[->, shorten >=2pt] (LDAmid)  -- (LDAT);
        \draw[->, shorten >=2pt] (LDAT)  -- (mis);
        \draw[->, shorten >=2pt] (corr)  -- (mis);
        \draw[->, shorten >=2pt] (hierarchy)  -- (mis);
    }
    \caption{MIS-2 Architecture}
\end{figure}

\vspace{-0.3cm}
Thus, we have significantly improved upon the MIS-1 architecture by addressing the following issues : 
\vspace{-0.3cm}
\begin{itemize}
    \item We infer the number of MIS-industries $K$ from the data by utilizing Bayesian Non-Parametrics. 
    \item We mitigate numerical instability by utilizing an ensemble to get an Empirical Bayes prior. 
    \item We temporally evolve our MIS-industries by utilizing a Markov parameter process. 
    \item We directly account for correlated/hierarchical topics (see next section). 
\end{itemize}
\vspace{-0.3cm}
We have found this architecture to be highly numerically stable and robust to small amounts of noise in the data (assuming the data has been appropriately pre-processed, as discussed in Section 2). Though this is an elaborate network of models, each component can be analyzed independently, making it highly human-interpretable and allowing for robust model risk management in practice. 

An obvious limitation of the MIS-2 architecture is that once we set $K$ in $\text{year}=1$ it is fixed for all following years, which is something we hope to improve upon in a future paper. In theory, a simple solution could be to use a Markov sequence of HDP models rather than LDA models, however we found this to be numerically unstable in practice and thus unreliable for our application.

\newpage
\section{Industry-Mixture Post-Processing}

Though we do not use CTM and HLDA to model MIS-industry relationships directly, we instead introduce two post-processing transformations on the posterior for $\boldsymbol\theta$ that capture the spirit of those architectures. As with our text pre-processor, these adjustments must be defined by a practitioner with sufficient domain expertise, such that all links are comprehensively defined and the model is internally consistent. Similar to semantic trees in Section 2, we construct \textbf{MIS-industry networks} to represent sub/super relations (arrow) and correlations (dashed line), such as in the following illustrative example. Each $k$-th MIS-industry is named after the keyphrase in $\boldsymbol\phi_k$ with highest probability : 
\begin{figure}[h!]
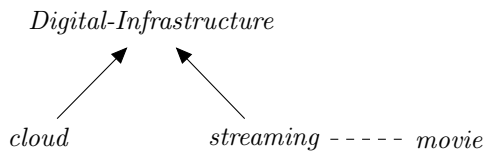

    \centering
    \tikz{
        \node (digital_infrastructure) {\textit{Digital-Infrastructure}};
        \node [below=of digital_infrastructure, xshift=-1.5cm] (cloud) {\textit{cloud}};
        \node [below=of digital_infrastructure, xshift=+1.5cm] (streaming) {\textit{streaming}};
        \node [right=of streaming] (movie) {\textit{movie}};
        \draw[->, shorten >=2pt] (cloud)  -- (digital_infrastructure);
        \draw[->, shorten >=2pt] (streaming)  -- (digital_infrastructure);
        \draw[dashed] (streaming) -- (movie)
    }
    \caption{Example of MIS-Industry Network}
\end{figure}

\vspace{-0.3cm}

We can then use this MIS-industry network to make adjustments to each $\boldsymbol\theta_m$ based on a simple rule-set, producing \emph{improper simplices} (discrete distributions over categories that sum to something greater than one). This isn't any sort of issue in practice, but changes the way we think of each element of $\boldsymbol\theta_m$ to be an \textbf{MIS-industry relevance score} rather than a part of a mixture. We illustrate the two adjustment rules with simple examples, though they can be easily generalized. 

\textbf{Correlation Adjustment :} Suppose we have three MIS-industries $\{A,B,C\}$ where $A$ and $B$ are correlated. We then transform the raw industry-mixture $\boldsymbol\theta_m = [0.1, 0.3, 0.6]$ into $\boldsymbol\theta_m^* = [0.4, 0.4, 0.6]$ by : 
\vspace{-0.3cm}
\begin{itemize}
    \item $\mathbb P(A\mid \text{Firm}_m) \leftarrow \mathbb P(A \mid \text{Firm}_m) + \mathbb P(B \mid \text{Firm}_m) = 0.1 + 0.3 = 0.4$
    \item $\mathbb P(B\mid \text{Firm}_m) \leftarrow \mathbb P(B \mid \text{Firm}_m) + \mathbb P(A \mid \text{Firm}_m) = 0.3 + 0.1 = 0.4$
\end{itemize}
\vspace{-0.3cm}

\textbf{Hierarchy Adjustment :} Similarly, suppose we have four MIS-industries $\{a_1, a_2,A,B\}$ where $a_1$ and $a_2$ are each sub-industries of $A$. We then transform the raw industry-mixture $\boldsymbol\theta_m = [0.1, 0.2, 0.3, 0.4]$ into $\boldsymbol\theta_m^* = [0.4, 0.5, 0.6, 0.4]$ by : 
\vspace{-0.3cm}
\begin{itemize}
    \item $\mathbb P(a_1\mid \text{Firm}_m) \leftarrow \mathbb P(a_1 \mid \text{Firm}_m) + \mathbb P(A \mid \text{Firm}_m) = 0.1 + 0.3 = 0.4$
    \item $\mathbb P(a_2\mid \text{Firm}_m) \leftarrow \mathbb P(a_2 \mid \text{Firm}_m) + \mathbb P(A \mid \text{Firm}_m) = 0.2 + 0.3 = 0.5$
    \item $\mathbb P(A\mid \text{Firm}_m) \leftarrow \mathbb P(A \mid \text{Firm}_m) + \mathbb P(a_1 \mid \text{Firm}_m) + \mathbb P(a_2 \mid \text{Firm}_m) = 0.3 + 0.1 + 0.2 = 0.6$
\end{itemize}
\vspace{-0.3cm}
In our implementation of MIS-2, we have several hundred links between MIS-industries, which allows us to produce ``baseball cards'' that summarize a firm as follows :

\begin{figure}[h!]
    \centering
    \includegraphics[width=11cm]{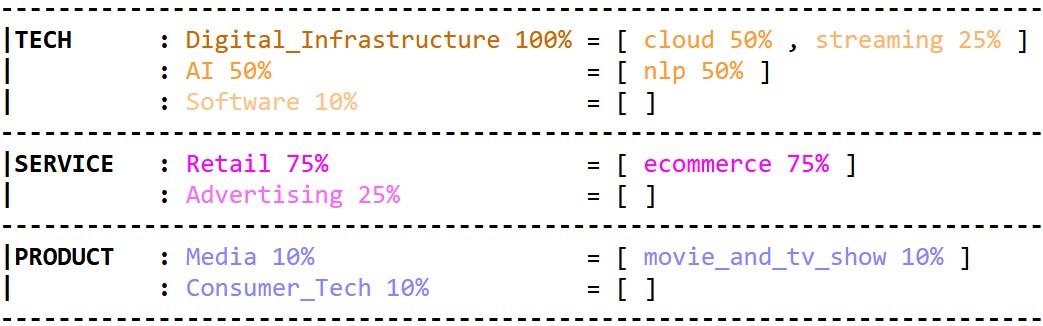}
    \caption{MIS-Industry Relevance Scores for Amazon}
\end{figure}
Here we highlight just how expressive MIS-2 can be, which summarizes much more information for Amazon than GICS is able to do. We'll now demonstrate several particular portfolio management applications in which we leverage MIS-2 to solve practical problems. 

\newpage
\section{Applications}
\subsection{Sector / Industry / Thematic Portfolios}

A popular style of active investing involves concentrating a bet around a particular product or service. At the broadest level, \textbf{sector portfolios} offer exposure to a wide category such as \emph{information technology} to capture potential upside associated with technological innovation. Alternatively, at a niche level, an investor can construct a portfolio around a specific technology, such as robotics, which is often referred to as \textbf{thematic investing}. Generally speaking, sector and thematic investing are essentially the same idea - though sector portfolios almost \emph{always} rely on GICS, while thematic portfolios almost \emph{never} rely on GICS.  

Regardless of how you want to label such portfolios, MIS-2 offers a very simple and interpretable framework to construct portfolios focused on a specific product or service. Consider the following \emph{automation} thematic portfolio. We apologize for the small font, we encourage zooming in on a computer to read the text. 
\begin{figure}[h!]
    \centering
    \includegraphics[width=16.5cm]{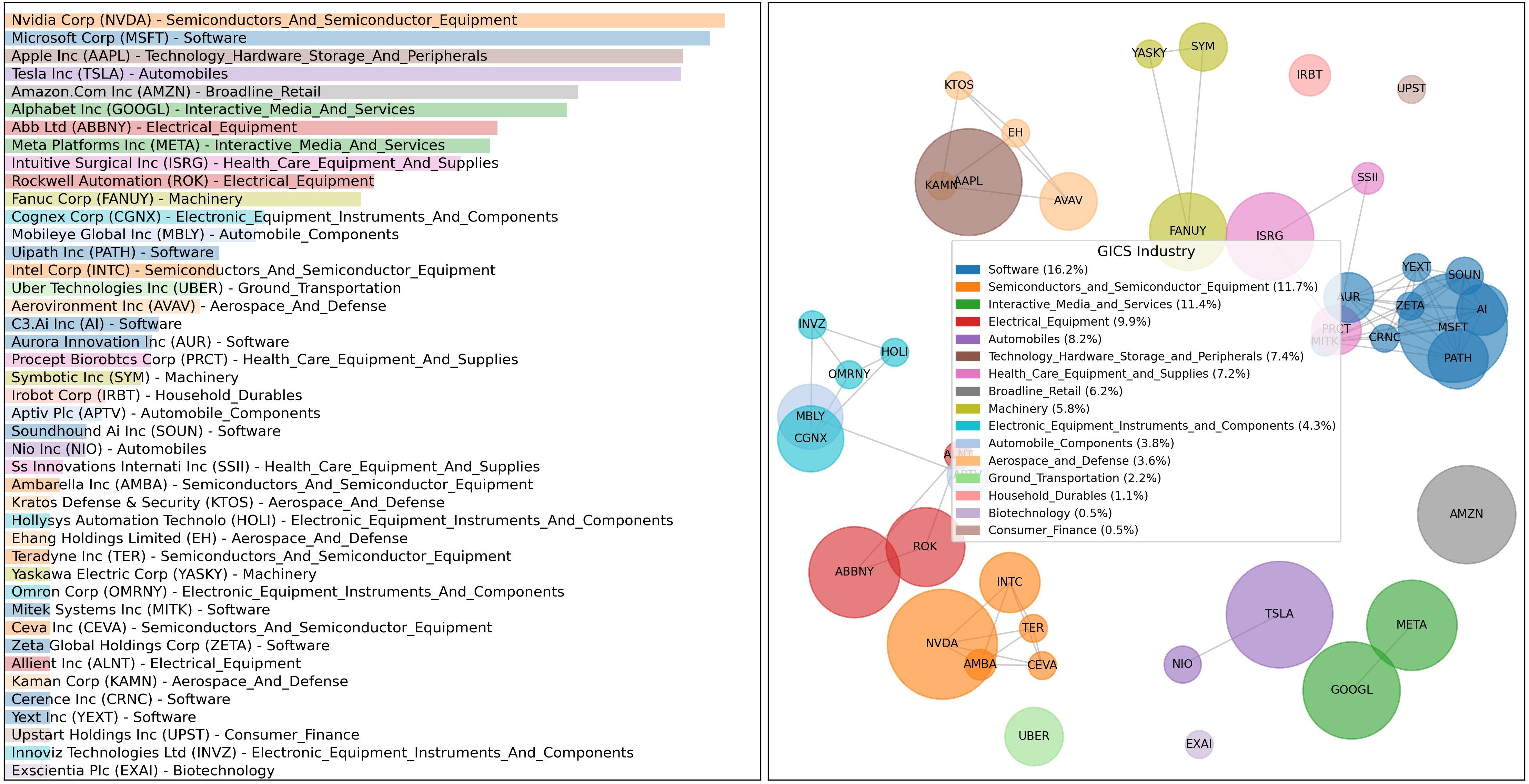}
    \caption{MIS-2 Thematic Portfolio : Automation}
\end{figure}

\vspace{-0.3cm}
This portfolio is here strictly for illustrative purposes, though the methodological framework can be easily generalized. We utilize the following procedure : 
\vspace{-0.3cm}
\begin{enumerate}
    \item We obtain MIS-industry relevance scores for all publicly traded firms. 
    \item We define a firm's \textbf{dollar exposure} to a theme as a firm's market cap multiplied by its relevance score to that theme (e.g. a $\$100$bn firm with $50\%$ relevance to \emph{automation} has $\$50$bn dollar exposure).
    \item We sort firms based on dollar exposure to a theme, and select the top 50 firms. 
    \item We perform mean-variance optimization with a risk model to obtain risk-managed weights. 
\end{enumerate}
\vspace{-0.3cm}
Since we use MIS-2 to construct our relevance signal, we're able to leverage the rich text of business descriptions to find many automation-oriented firms across many industries. You'll notice that this portfolio includes both hardware and software firms across both artificial intelligence and robotics. This type of portfolio construction is simply not possible in GICS, because GICS is a rigid system with much fewer industry labels than can be constructed with MIS-2. 

\newpage
\subsection{Nearest Neighbor Portfolios and Exclusion Lists}

A more complex task than constructing a single-industry portfolio is constructing a multi-industry portfolio. Consider an early investor in Amazon with a highly concentrated investment position in Amazon stock that they would like to wind down. This investor would like to maintain similar exposure to innovative tech firms, while diversifying across more stocks - thus, they would naturally need to know Amazon's \emph{nearest neighbors}. 

Identifying the 50 most similar firms to Amazon is a tricky task to handle - as Amazon is a multi-industry firm (see Figure 10). We offer a simple yet rigorous solution - define a distance metric on the space of MIS industry-mixtures. For firms $i$ and $j$, we define \textbf{text similarity} as the following overlap score : 
\begin{align}
    \rho^\text{text}_{i,j} = \text{similarity}\big(\mathbf x_i, \mathbf x_j\big) = \sum_{k=1}^K \min\big(\boldsymbol\theta_{i,k}, \boldsymbol\theta_{j,k}\big)
\end{align}
To make this similarity score even more robust, we can incorporate additional information, such as historical returns correlation and factor similarity (based on a risk model). Further, we can apply various penalties to filter out less-reliable neighbors, on the basis of things like  firm size or idiosyncratic risk - though we leave such implementation details to the discretion of the practitioner. We can thus define a \textbf{composite similarity score}, with $\lambda_1 + \lambda_2 + \lambda_3 = 1$, as : 
\begin{align}
    \rho^\text{composite}_{i,j} = \big(\lambda_1 \cdot \rho^\text{text}_{i,j}\big) + \big(\lambda_2 \cdot \rho^\text{returns}_{i,j}\big) + \big(\lambda_3 \cdot \rho^\text{factors}_{i,j}\big)
\end{align}
Utilizing a similar process to thematic portfolios, only replacing the relevance score with a composite similarity score, we obtain the following \textbf{nearest neighbor portfolio} for Amazon : 
\begin{figure}[h!]
    \centering
    \includegraphics[width=16.5cm]{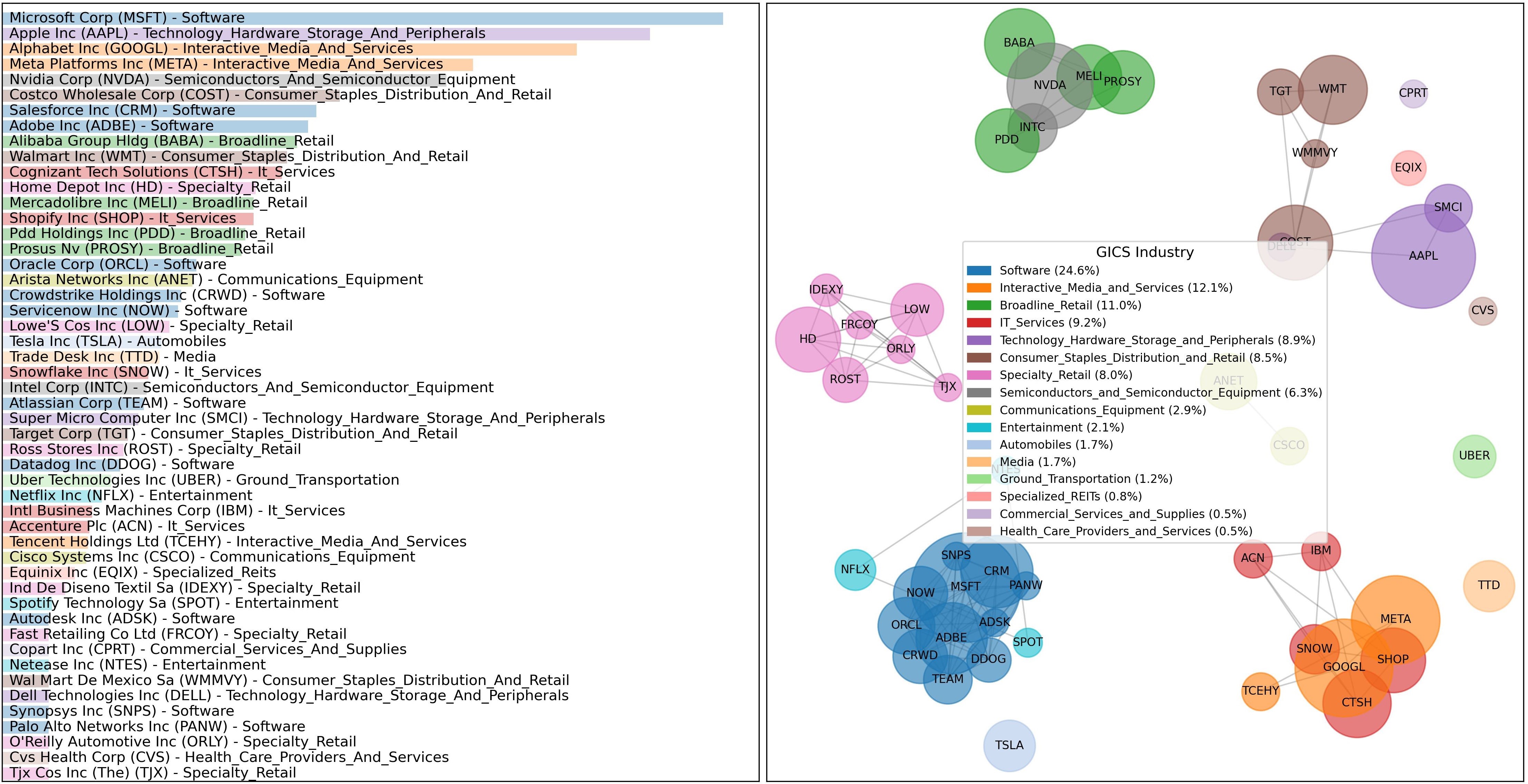}
    \caption{MIS-2 Nearest Neighbor Portfolio for Amazon}
\end{figure}

Here, you'll notice that Amazon has neighbors across multiple industries - which appropriately represents the inherent diversification of the firm's business ventures. Alternatively, investors can use this as an exclusion list rather than a portfolio if they want to specifically \emph{avoid} firms similar to Amazon, though the nearest neighbor list would be the same. Again, something like this wouldn't even be remotely possible with GICS, highlighting the benefit of our multi-industry approach. 

\newpage
\section{GICS vs MIS-2 : Out-of-Sample Testing}
\vspace{-0.3cm}
While up to this point we have only provided anecdotal reasoning to demonstrate the value of MIS-2, we will now rigorously argue that MIS-2 is a superior \emph{risk management} tool. 

A primary reason for the utilization of sector and industry information in asset management is the fact that firms involved in similar products and services tend to have correlated returns. When a firm is compared to its peers, this is often on the basis of industry - thus, in order to compare two industry classification systems we will compare GICS and MIS in terms of their ability to predict future returns correlations. 

We will conduct our test as follows : 
\vspace{-0.3cm}
\begin{itemize}
    \item Per firm, we construct two peer portfolios : an MIS-Neighborhood portfolio, which is the 50 nearest neighbors of a firm based on our composite similarity score, and a GICS-industry portfolio based on that firm's GICS industry. Both portfolios exclude the firm itself. 
    \item Next, we get future 1-year returns correlations between a firm and each of its two corresponding peer portfolios. A higher value is better, as it indicates more-relevant peers. 
    \item Finally, we take the difference between the MIS-correlation and the GICS-correlation. If the difference is positive, then the MIS peers are a better predictor of a firm's returns than its GICS peers. 
\end{itemize}
\vspace{-0.3cm}
We perform this test for each firm in our universe of over 3000 stocks, and summarize the data below, grouped by GICS sector. We train up until 2022, and test in 2023 so that we have out-of-sample results : 
\begin{figure}[h!]
    \centering
    \includegraphics[width=15cm]{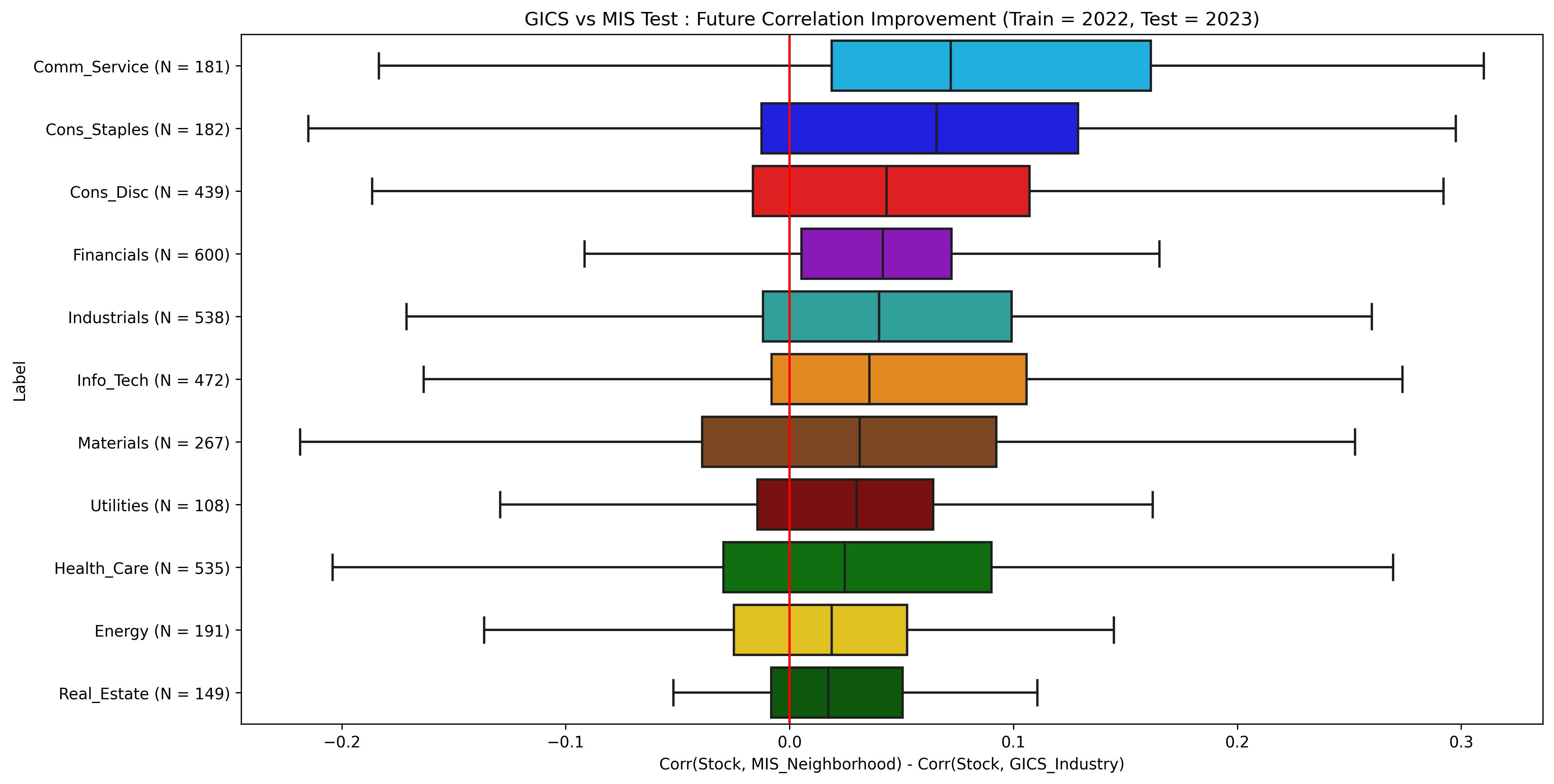}
    \caption{Out-of-Sample Test, by GICS Sector}
\end{figure}

\vspace{-0.3cm}
You'll notice that on average, MIS-2 outperforms GICS consistently across all sectors. There are of some instances in which GICS is better - though these are not common. Firms in Communication Services tend to get the highest boost from MIS-2, as many large tech firms tend to involve themselves in many sectors and thus are better represented by a multi-industry model. 

For our particular implementation, we only have data from 2021 to present (due to data budget constraints), and thus we are not able to perform a longer backtest - though we eagerly hope to address this in future iterations.  For the reader who is interested in an even more granular comparison, we decompose the above data on the next page in a figure that partitions firms by GICS industry rather than by GICS sector, sorted from highest-to-lowest median. Again, we apologize for the small font, though hopefully it is easy to zoom in on a computer. It tells the same story, though in more detail. 

\newpage

\begin{figure}[h!]
    \centering
    \includegraphics[width=15cm]{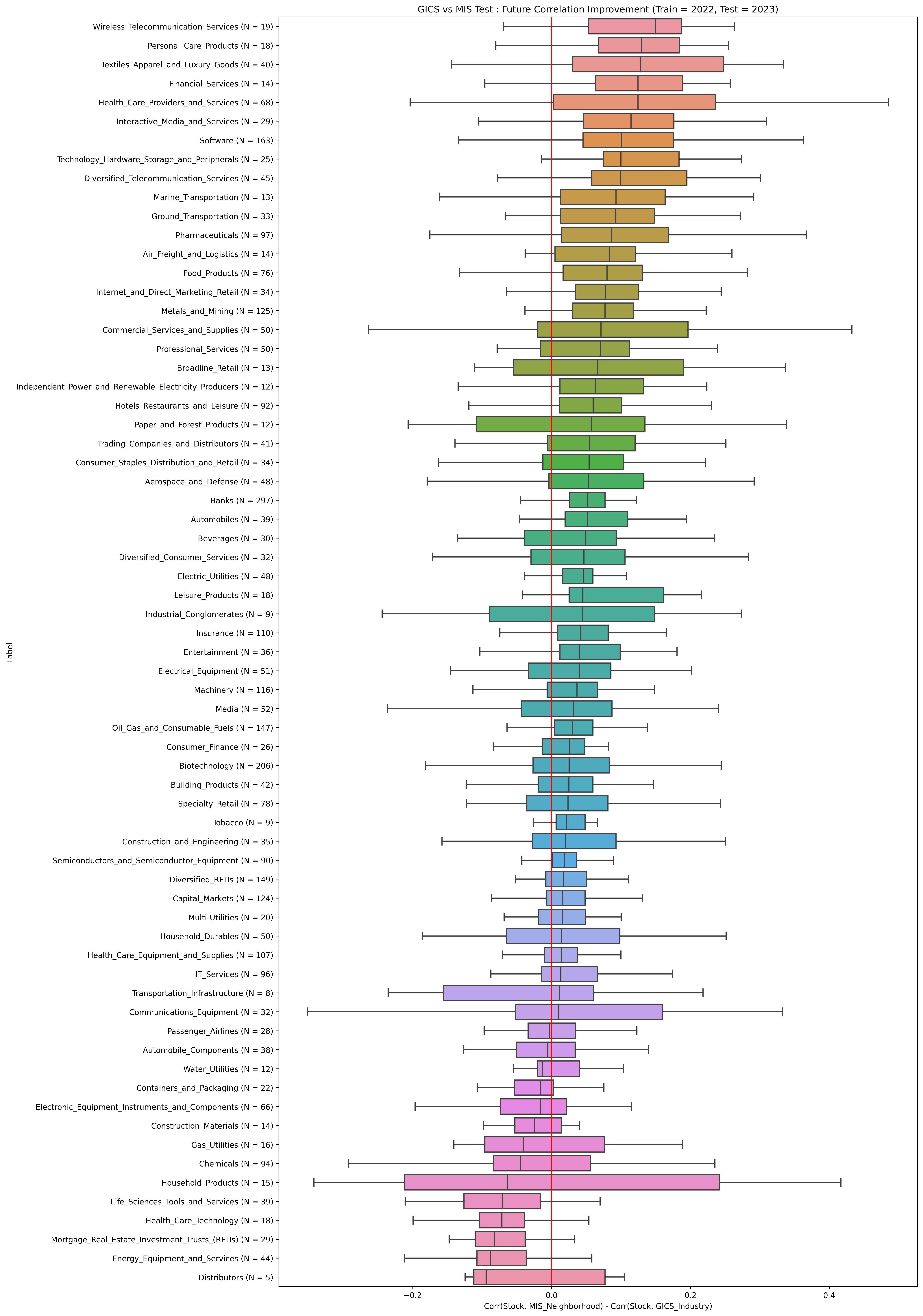}
    \caption{Out-of-Sample Test, by GICS Industry}
\end{figure}

\newpage
\section{Conclusion}
\vspace{-0.4cm}
What we presented here is a middle-step in an ongoing process of research and development. MIS-1 had many critical limitations that we addressed in MIS-2, though there still exist other areas for potential improvement that we would like to iterate upon. We hope that we have presented a compelling argument for MIS-2 as an improved alternative to GICS, as a multi-industry classification model will allow asset managers to better identify and manage risk. 
\vspace{-0.7cm}
\bibliographystyle{ieeetr}
\bibliography{main} 

\end{document}